\documentclass[conference]{IEEEtran}
\IEEEoverridecommandlockouts

\usepackage{cite}
\usepackage{amsmath,amssymb,amsfonts}
\usepackage{algorithmic}
\usepackage{graphicx}
\usepackage{textcomp}
\usepackage{xcolor}
\usepackage{float} 
\usepackage{listings}
\usepackage[T1]{fontenc}
\usepackage[utf8]{inputenc}
\usepackage{url}
\usepackage{booktabs}
\usepackage{fancyhdr}
\usepackage{hyperref}

\lstdefinelanguage{json}{
    basicstyle=\ttfamily\footnotesize,
    numbers=left,
    numberstyle=\tiny\color{gray},
    stepnumber=1,
    numbersep=5pt,
    showstringspaces=false,
    breaklines=true,
    frame=tb,
    backgroundcolor=\color{gray!10},
    literate=
     *{0}{{{\color{blue}0}}}{1}
      {1}{{{\color{blue}1}}}{1}
      {2}{{{\color{blue}2}}}{1}
      {3}{{{\color{blue}3}}}{1}
      {4}{{{\color{blue}4}}}{1}
      {5}{{{\color{blue}5}}}{1}
      {6}{{{\color{blue}6}}}{1}
      {7}{{{\color{blue}7}}}{1}
      {8}{{{\color{blue}8}}}{1}
      {9}{{{\color{blue}9}}}{1}
      {:}{{{\color{red}:}}}{1}
      {,}{{{\color{red},}}}{1}
      {\{}{{{\color{red}\{}}}{1}
      {\}}{{{\color{red}\}}}}{1}
      {[}{{{\color{red}[}}}{1}
      {]}{{{\color{red}]}}}{1},
}
\begin{document}

\title{Development and Application of a Decentralized Domain Name Service}

\author{\IEEEauthorblockN{Guang Yang}
\IEEEauthorblockA{University of California, Berkeley\\
guangyang19@berkeley.edu
\\version 0.2}
}

\maketitle

\begin{abstract}
The current Domain Name System (DNS), as a core infrastructure of the internet, exhibits several shortcomings: its centralized architecture leads to censorship risks and single points of failure, making domain name resolution vulnerable to attacks. The lack of encryption in the resolution process exposes it to DNS hijacking and cache poisoning attacks~\cite{dnssec_icann}. Additionally, the high operational costs limit participation and innovation among small to medium-sized users~\cite{vakali2003content}. To address these issues, this paper proposes a Decentralized Domain Name Service (DDNS) based on blockchain (Phicoin~\cite{phicoin_whitepaper}) and distributed storage (IPFS). By leveraging the immutability of blockchain and the content verification of IPFS~\cite{ipfs_docs}, the system achieves decentralized storage and distribution of domain name records, eliminating the centralized dependencies of traditional DNS. With a block time of 15 seconds, the system supports rapid broadcasting of domain name updates, significantly improving resolution efficiency. The DDNS aims to serve as a complement or backup to the existing DNS system, providing a pollution-resistant, censorship-resistant, high-performance, and low-cost domain name resolution solution, offering a new technical path for the security and stability of the internet.
\end{abstract}


\section{Introduction}

\subsection{Background}

The Domain Name System (DNS) is a critical infrastructure of the internet, responsible for translating user-friendly domain names into machine-readable IP addresses~\cite{mockapetris1987rfc1034}. It serves as the backbone of internet communication, enabling users to access websites and online services seamlessly. However, the traditional DNS system, due to its centralized architecture, has gradually exposed several limitations~\cite{gao2013empirical}:

\begin{itemize}
    \item \textbf{Single Point of Failure:} Centralized authoritative and caching servers can lead to widespread service interruptions if they fail~\cite{moura2016anycast}.
    \item \textbf{Susceptibility to Censorship:} The centralized architecture makes DNS services prone to control or interference, allowing manipulation of resolution records to restrict user access~\cite{kalodner2015empirical}.
    \item \textbf{High Operational Costs:} The substantial deployment and maintenance costs of authoritative and caching servers limit participation by small and medium-sized enterprises and individual users~\cite{vakali2003content}.
\end{itemize}

Common threats to the DNS system include:

\begin{itemize}
    \item \textbf{DNS Hijacking:} Attackers tamper with resolution records to redirect legitimate domain names to malicious IP addresses, luring users to phishing sites or ad pages~\cite{li2021bdns}.
    \item \textbf{DNS Cache Poisoning:} Injection of forged resolution results into DNS server caches, causing users to be redirected to incorrect addresses~\cite{son2010hitchhiker}.
    \item \textbf{Censorship and Blocking:} Modification or blocking of resolution requests to restrict access to specific domain names or content~\cite{patsakis2020unravelling}.
\end{itemize}

\subsection{Motivation}

With the continuous growth of the internet, the centralized architecture of the traditional DNS system faces increasing challenges in security, reliability, and scalability~\cite{herzberg2013dnssec}. Centralized authoritative servers not only become prime targets for network attacks and censorship but also affect the availability of domain name resolution services due to single points of failure~\cite{moura2016anycast}. Furthermore, attacks like DNS hijacking and cache poisoning exploit the lack of data integrity verification in traditional DNS systems, leading to tampered resolution records and severely threatening user security and privacy~\cite{liu2018high}. These issues highlight the urgent need for a technical solution that can compensate for the shortcomings of the traditional DNS system.

Decentralized technologies offer a new approach to domain name resolution systems~\cite{liu2018high}. By utilizing the distributed ledger and immutability of blockchain~\cite{nakamoto2008bitcoin}, domain name records can be managed in a decentralized manner, avoiding single points of failure and data tampering. Combining this with IPFS's distributed storage capabilities~\cite{benet2014ipfs}, the storage and distribution of resolution records become more flexible and efficient, capable of meeting future complex DNS protocol extension needs. This paper aims to develop a Decentralized Domain Name Service (DDNS) that, through a decentralized architecture and cryptographic verification mechanisms, achieves pollution resistance, censorship resistance, high performance, and low cost in domain name resolution services, providing a reliable complement or alternative to the existing DNS system.

\subsection{Related Work}

Existing decentralized DNS solutions like Ethereum Name Service (ENS)~\cite{ens_docs} and Namecoin~\cite{kalodner2015empirical} have made significant strides in enhancing privacy and censorship resistance.

\begin{itemize}
    \item \textbf{ENS:} Utilizes the Ethereum blockchain to manage domain names ending with \texttt{.eth}, allowing users to associate metadata and cryptocurrency addresses with human-readable names~\cite{xia2021ens}.
    \item \textbf{Namecoin:} A fork of Bitcoin that enables the registration of \texttt{.bit} domains on its blockchain~\cite{kalodner2015empirical}.
\end{itemize}

\textbf{Advantages of Existing Solutions:}

\begin{itemize}
    \item Enhanced privacy.
    \item Resistance to censorship.
    \item Elimination of central authority control.
\end{itemize}

\textbf{Limitations of Existing Solutions:}

\begin{itemize}
    \item Performance issues due to blockchain scalability constraints~\cite{carlsten2016instability}.
    \item Potential centralization risks in the management of top-level domains~\cite{patsakis2020unravelling}.
    \item Higher costs associated with blockchain transactions~\cite{ali2016blockstack}.
\end{itemize}

\textbf{Our Contribution:}

The DDNS system proposed in this paper introduces the Phicoin blockchain~\cite{phicoin_whitepaper}, designed specifically for domain name services with a 15-second block time to enhance performance. The system focuses on fair participation, asset compatibility, and ultra-low costs, addressing the limitations of existing solutions by improving scalability and reducing operational expenses.

\begin{table}[H]
  \centering
  \caption{Comparison of Decentralized DNS Solutions}
  \resizebox{0.45\textwidth}{!}{%
  \begin{tabular}{lccccc}
  \toprule
  \textbf{Feature} & \textbf{Traditional DNS} & \textbf{ENS} & \textbf{Namecoin} & \textbf{Handshake} & \textbf{Phicoin (This Work)} \\
  \midrule
  Decentralization     & No              & Yes          & Yes          & Yes          & Yes              \\
  Performance          & High            & Medium       & Medium       & Medium       & High             \\
  Censorship Resistant & Low             & High         & High         & High         & High             \\
  Cost (per domain)    & High            & High         & Medium       & Medium       & Super Low (\$0.00025)  \\
  Blockchain Speed     & N/A             & 15 sec       & 10 min       & 10 min       & 15 sec           \\
  Extensibility        & Limited         & Flexible     & Limited      & Limited      & Flexible         \\
  \bottomrule
  \end{tabular}%
  }
  \end{table}
  
\section{System Design}

\subsection{Architecture Overview}

The DDNS system leverages blockchain and distributed storage technologies to decentralize domain name resolution. The key components include:

\begin{itemize}
    \item \textbf{Phicoin Blockchain:} Used for domain asset binding and verification. It records domain name ownership and updates, ensuring immutability and transparency~\cite{phicoin_whitepaper}.
    \item \textbf{IPFS (InterPlanetary File System):} Used for distributed storage of resolution data. It stores domain resolution data (e.g., IP addresses), enabling decentralized retrieval of this information~\cite{benet2014ipfs}.
\end{itemize}

\begin{figure}[H]
\centering
\includegraphics[width=0.35\textwidth]{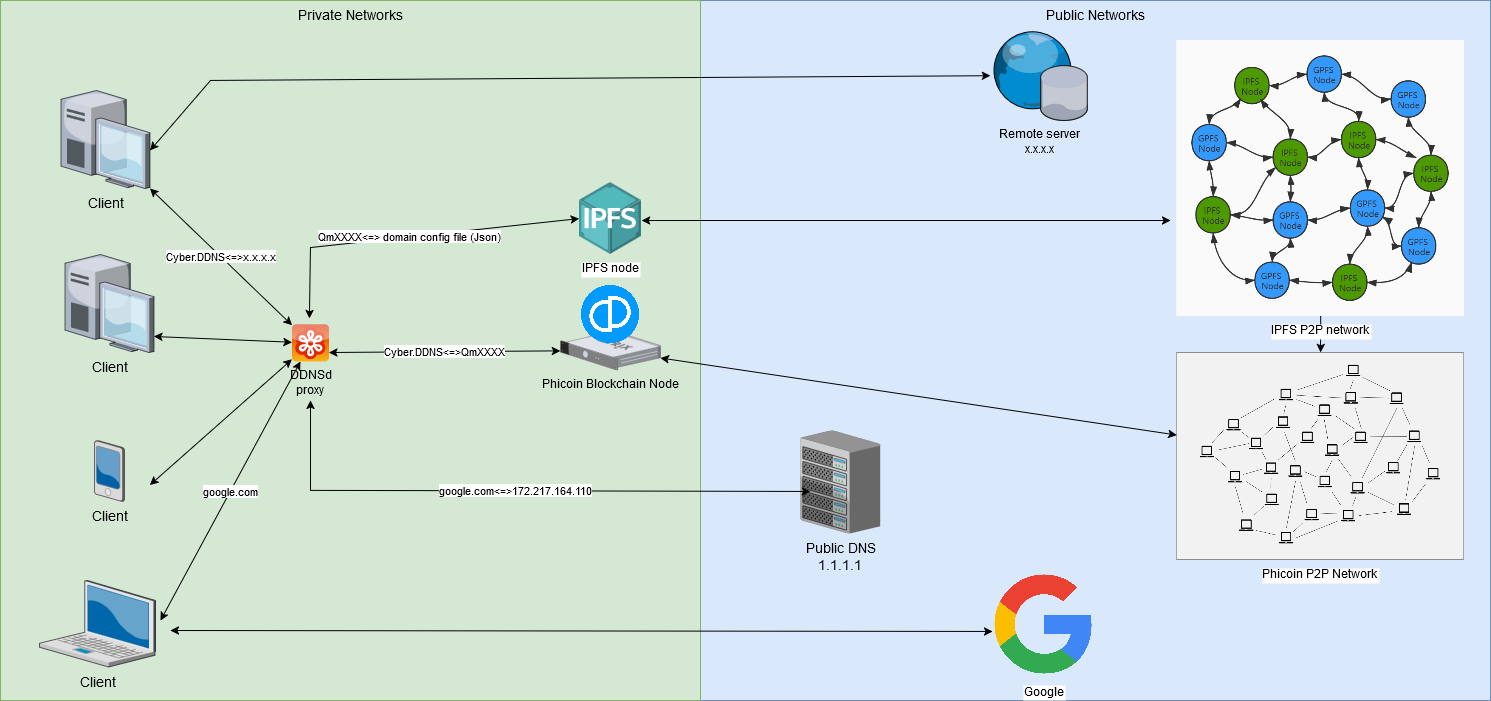}
\caption{DDNS System Architecture}
\end{figure}

\subsection{Key Components}

\subsubsection{Phicoin Blockchain}

\textbf{Design Goals:}

\begin{enumerate}
    \item \textbf{Economic Efficiency:} Achieve ultra-low costs to encourage widespread adoption~\cite{phicoin_whitepaper}.
    \item \textbf{Fair Participation:} Ensure equal opportunities for all participants without central nodes or masternodes~\cite{phicoin_whitepaper}.
    \item \textbf{Asset Compatibility:} Support domain name assets and other digital assets~\cite{phicoin_whitepaper}.
\end{enumerate}

\textbf{Current Participation:}

\begin{itemize}
    \item \textbf{Test Mining Phase:} Over 1,800 participants.
    \item \textbf{Post-Mainnet Launch:} Over 300 global nodes running.
\end{itemize}

Phicoin utilizes the ASIC-resistant \textbf{Phihash} algorithm, mined using GPUs~\cite{phicoin_whitepaper}. The value is determined by the user's work contribution and recognition, promoting true decentralization through self-incentivization.

\subsubsection{Decentralized Domain Name Service Protocol}

\textbf{Domain Types:}

\begin{table}[h!]
    \centering
    \caption{Phicoin Domain Rules and Properties}
    \scalebox{0.7}{
      \begin{tabular}{@{}ll@{}}
        \toprule
        \textbf{Category}                 & \textbf{Description}                                                                                   \\ \midrule
        \textbf{Phicoin Top-Level Domains} & Used to create Phicoin top-level domains (pTLDs), such as \texttt{.ddns}.                              \\
                                           & Anyone with a GPU can mine Phicoin and create a TLD~\cite{phicoin_whitepaper}.                        \\ \midrule
        \textbf{Sub Domains}              & Used to create second-level domains with specific rules:                                              \\
        \midrule
        \textbf{Asset Structure}          & Maximum of 32 characters.                                                                             \\
        \textbf{Data Structure}           & \texttt{!} denotes root asset, \texttt{/} denotes separator.                                           \\
        \textbf{Properties}               & \begin{tabular}[t]{@{}l@{}}
                                             \textbullet\ Non-reissuable. \\
                                             \textbullet\ Quantity of 1. \\
                                             \textbullet\ Unit of 1.
                                           \end{tabular}                                                                                          \\ 
        \textbf{Initial Binding Hash}     & \texttt{000\ldots000} (64 zeros).                                                                     \\
        \textbf{Deactivation Hash}        & \texttt{Qm000\ldots000} (46-character IPFS hash).                                                     \\
        \textbf{Management Rights}        & Transferred to a specified address upon creation.                                                     \\
        \textbf{Fees}                     & \begin{tabular}[t]{@{}l@{}}
                                             \textbullet\ Creation Fee: 0.1 phi. \\
                                             \textbullet\ Modification Fee: 0.1 phi.
                                           \end{tabular}                                                                                          \\
        \textbf{No Annual Fees}           & Domains never expire.                                                                                 \\
        \textbf{Subdomains}               & Supports creation of subdomains up to a total length of 30 characters.                                \\ \bottomrule
      \end{tabular}
    }
    
    \label{tab:phicoin_domains}
  \end{table}

\subsubsection{Domain Templates}

Domain records are defined using JSON files, enabling flexibility and extensibility~\cite{ens_docs}.

\textbf{Basic Record Types:}

\begin{itemize}
    \item \textbf{Type A (IPv4 Address):}

\begin{lstlisting}[language=json]
{
  "Type": "A",
  "Address": "192.168.1.1"
}
\end{lstlisting}

    \item \textbf{Type AAAA (IPv6 Address):}

\begin{lstlisting}[language=json]
{
  "Type": "AAAA",
  "Address": "2001:db8::1"
}
\end{lstlisting}

    \item \textbf{Type CNAME (Canonical Name):}

\begin{lstlisting}[language=json]
{
  "Type": "CNAME",
  "Target": "example.com"
}
\end{lstlisting}

    \item \textbf{Type MX (Mail Exchange Record):}

\begin{lstlisting}[language=json]
{
  "Type": "MX",
  "MailServer": "mail.example.com",
  "TTL": 3600,
  "Priority": 10
}
\end{lstlisting}

\end{itemize}

\textbf{Extensibility:}

Users can extend these templates to include additional record types as needed, such as TLSA for DANE (DNS-based Authentication of Named Entities) records~\cite{hoffman2012dns}.

\subsubsection{IPFS Integration}

\textbf{Binding IPFS Hashes with Domain Assets:}

\begin{itemize}
    \item Users create and host the domain configuration JSON files using an IPFS client~\cite{benet2014ipfs}.
    \item The data is distributed across the IPFS network.
    \item In this study, Pinata's free API is utilized, allowing management of up to 500 files—sufficient for typical domain resolution needs.
\end{itemize}

\subsubsection{Domain Resolution}

\textbf{Design Goals:}

\begin{enumerate}
    \item \textbf{Compatibility with Existing DNS:} Ensure that traditional DNS queries are handled appropriately.
    \item \textbf{DDNS Protocol Domain Resolution:} Enable resolution of DDNS domains.
\end{enumerate}

\textbf{Implementation:}

\begin{itemize}
    \item \textbf{Local DNS Proxy Service Deployment:}
    \begin{itemize}
        \item \textbf{Traditional Domain Queries:} Forwarded to public DNS servers like 1.1.1.1.
        \item \textbf{DDNS Domain Queries:}
        \begin{enumerate}
            \item Retrieve the IPFS hash from the Phicoin blockchain based on the domain name~\cite{phicoin_whitepaper}.
            \item Fetch the domain configuration JSON file from IPFS using the hash.
            \item Parse the JSON file to extract the domain records.
            \item Return the resolution result to the user.
        \end{enumerate}
    \end{itemize}
\end{itemize}

\begin{figure}[H]
\centering
\includegraphics[width=0.35\textwidth]{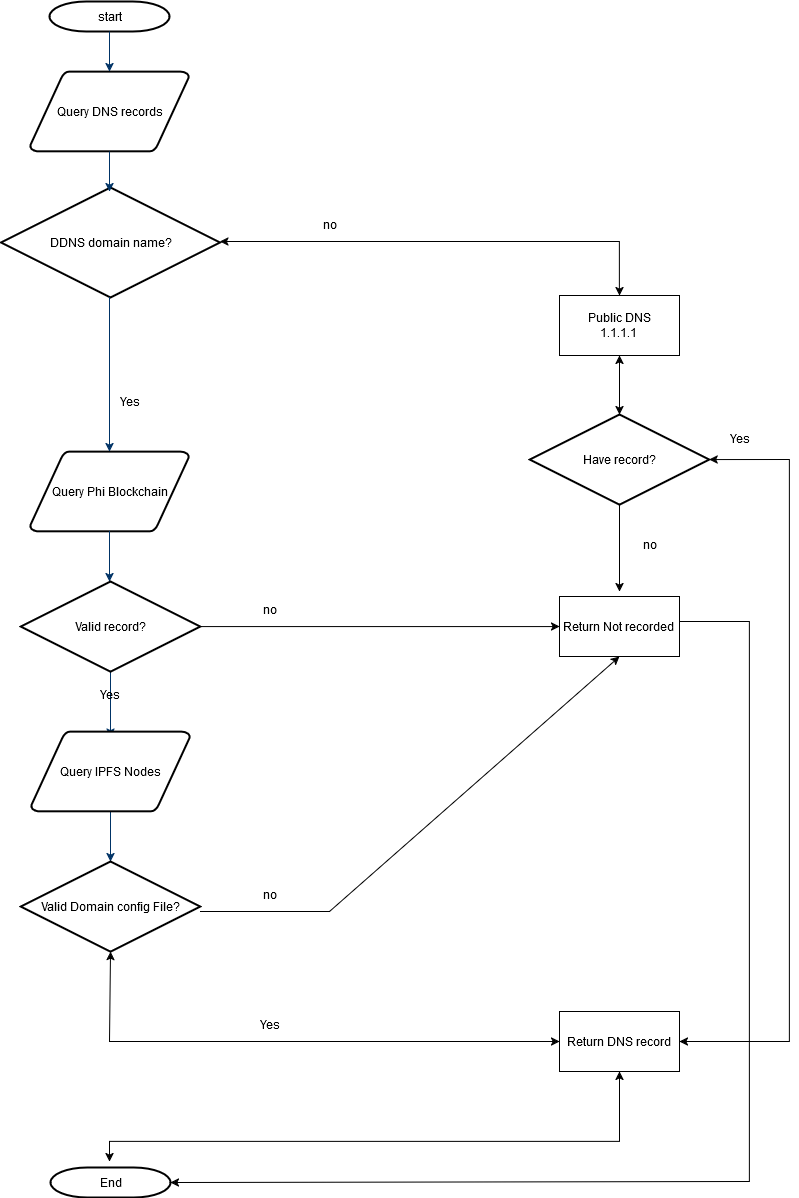}
\caption{DDNS Domain Resolution Process}
\end{figure}

\section{Implementation}

\subsection{Core Functions}

\begin{itemize}
    \item \textbf{Registering pTLDs:} Creation of new top-level domains on the Phicoin blockchain~\cite{phicoin_whitepaper}.
    \item \textbf{Adding Subdomains:} Generation of second-level and subdomains under a pTLD.
    \item \textbf{Disabling Subdomains:} Deactivation of subdomains by updating the associated IPFS hash.
    \item \textbf{Setting Domain Records:} Updating the domain's IPFS hash to point to the correct JSON configuration file.
    \item \textbf{DNS Domain Resolution Forwarding Service:} Handling traditional DNS queries by forwarding them to standard DNS servers.
    \item \textbf{DDNS Domain Resolution Service:} Resolving DDNS domains using the Phicoin blockchain and IPFS.
\end{itemize}

\subsection{Codebase and Modules}

\textbf{Main Modules:}

\begin{itemize}
    \item \texttt{ddnsd.py}: The primary daemon responsible for handling DNS queries. It interfaces with the Phicoin blockchain and IPFS to resolve domain names and return the results to clients.
    \item \texttt{ddns\_domain\_registration.ipynb}: A Jupyter Notebook that demonstrates how to register domains, manage subdomains, and update domain records on the blockchain.
\end{itemize}

\textbf{Module Functionalities:}

\begin{itemize}
    \item \textbf{Blockchain Interaction:} Functions to interact with the Phicoin blockchain for querying domain assets and associated IPFS hashes~\cite{phicoin_whitepaper}.
    \item \textbf{IPFS Interaction:} Methods to retrieve and pin JSON configuration files on IPFS.
    \item \textbf{DNS Server Integration:} Implements a DNS server that can handle both traditional DNS and DDNS queries.
    \item \textbf{Configuration Management:} Tools for domain owners to create and update domain records.
\end{itemize}

\section{Evaluation}
\section{Validation and Testing}
In order to validate our Decentralized Domain Name Service (DDNS), we performed an end-to-end experiment covering record storage on IPFS, on-chain registration via the Phicoin blockchain, and final DNS resolution through a local proxy server. This section details each step, including the core Python code, transaction outputs, and relevant screenshots.

\subsection{Overall Workflow}
The DDNS testing process involved the following:
\begin{enumerate}
    \item \textbf{Record Preparation:} Construct JSON data for different DNS record types (A, AAAA, CNAME, MX).
    \item \textbf{IPFS Upload:} Use Pinata’s API to store these JSON files, obtaining immutable content hashes.
    \item \textbf{Blockchain Update:} Broadcast each IPFS hash to the Phicoin network via \lstinline|set_domain_record|, anchoring the domain name and record hash on-chain.
    \item \textbf{DNS Proxy Query:} Run \texttt{nslookup} commands against a custom DNS proxy (port 5553) to confirm correct resolution.
\end{enumerate}

\subsection{Code Snippets and Transaction Outputs}
Below, we show two representative code segments. The first registers an \textbf{A record} (subdomain \texttt{www.xxx.ddns} $\to$ \texttt{1.2.3.4}); the second registers an \textbf{AAAA record} (\texttt{ipv6.xxx.ddns} $\to$ \texttt{2001:0000:130F:...}). Other record types (CNAME, MX) follow the same pattern.

A Record Registration

\begin{lstlisting}[language=Python, caption={Registering an A record for www.xxx.ddns}]
# 1) Construct the A record JSON
a_record = {
  "Type": "A",
  "Address": "1.2.3.4"
}

# 2) Upload to IPFS
subdomain = "www"
domain_name = f"{subdomain}.{test_top_domain}.{ROOT_DOMAIN}"  # => "www.xxx.ddns"
ipfs_hash = upload_to_pinata(domain_name.upper(), a_record)
print(f"IPFS Hash for {domain_name}: {ipfs_hash}")
# 3) Broadcast to Phicoin blockchain
update_result = set_domain_record(test_top_domain, subdomain, owner_address, ipfs_hash)
print(f"Domain Update Result: {update_result}")
\end{lstlisting}

\noindent
The \lstinline|update_result| dictionary typically includes a \texttt{result} field containing one or more transaction hashes. For instance:
\begin{center}
\texttt{\{"result": ["0a34d9200db66..."], "error": None, "id": "curltest"\}}
\end{center}
Such a hash indicates the on-chain confirmation that \texttt{www.xxx.ddns} is now bound to the specified IPFS record.

AAAA Record Registration
\begin{lstlisting}[language=Python, caption={Registering an AAAA record for ipv6.xxx.ddns}]
# 1) Construct the AAAA record JSON
aaaa_record = {
  "Type": "AAAA",
  "Address": "2001:0000:130F:0000:0000:09C0:876A:130B"
}

# 2) Upload to IPFS
subdomain = "ipv6"
domain_name = f"{subdomain}.{test_top_domain}.{ROOT_DOMAIN}"  # => "ipv6.xxx.ddns"
ipfs_hash = upload_to_pinata(domain_name.upper(), aaaa_record)
print(f"IPFS Hash for {domain_name}: {ipfs_hash}")
# 3) Broadcast to Phicoin blockchain
update_result = set_domain_record(test_top_domain, subdomain, owner_address, ipfs_hash)
print(f"Domain Update Result: {update_result}")
\end{lstlisting}

\begin{figure}[H]
    \centering
    \includegraphics[width=0.43\textwidth]{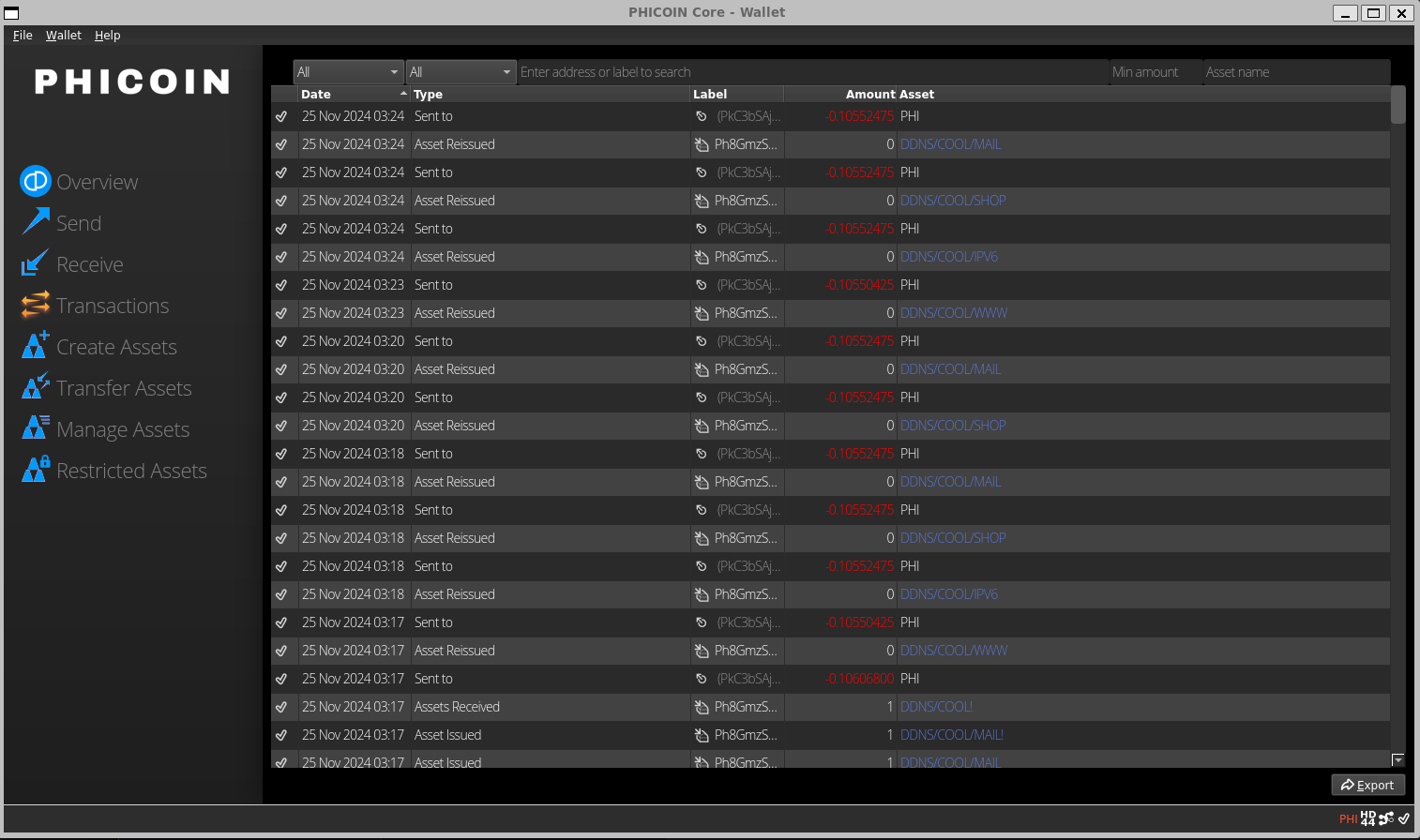}
    \caption{Phicoin QT client displaying finalized transactions for domain record updates.}
    \label{fig:phicoinqt}
\end{figure}

\noindent
Again, the \lstinline|update_result| indicates a successful transaction broadcast. Figure~\ref{fig:phicoinqt} shows the Phicoin QT client listing these transaction IDs, signifying that each domain record update has been mined.

\subsection{IPFS Verification}
\begin{figure}[H]
    \centering
    \includegraphics[width=0.43\textwidth]{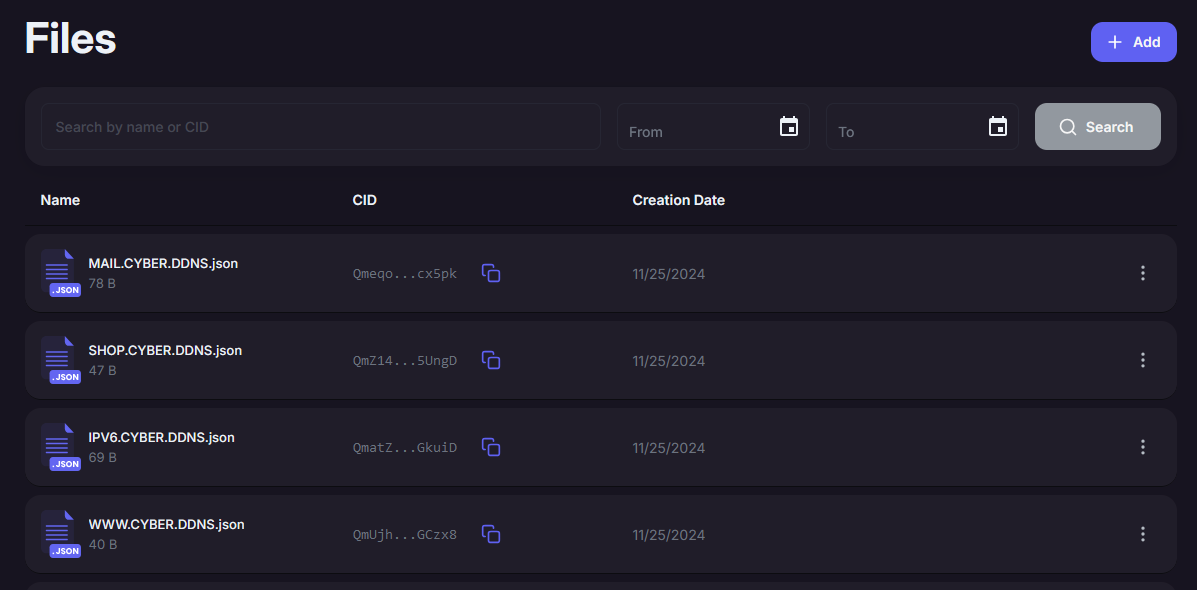}
    \caption{Pinata UI showing the newly uploaded JSON files mapped to unique IPFS hashes.}
    \label{fig:ipfs}
\end{figure}

After each upload, Pinata’s dashboard confirms the new content hash (Figure~\ref{fig:ipfs}). Storing records in IPFS ensures tamper-evident distribution, thus avoiding a single point of failure in the DNS data layer.

\subsection{DNS Proxy Resolution}
\begin{figure}[H]
    \centering
    \includegraphics[width=0.43\textwidth]{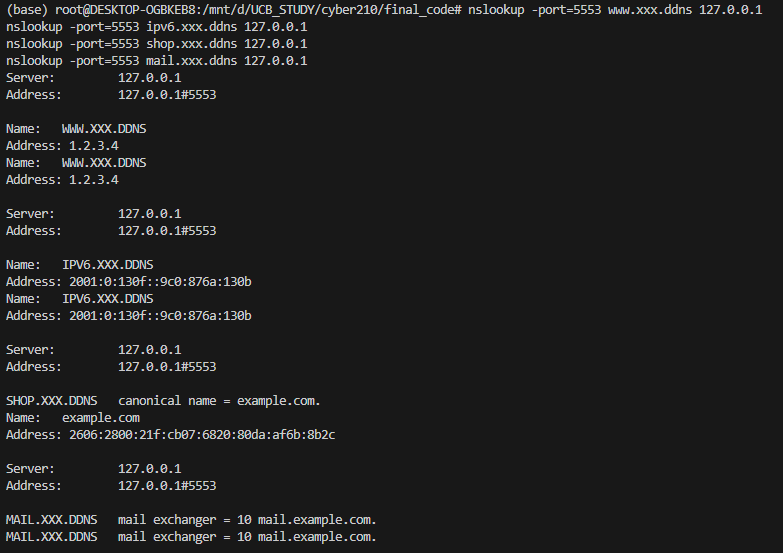}
    \caption{\texttt{nslookup} output demonstrating successful resolution of multiple record types.}
    \label{fig:nslookup}
\end{figure}

Finally, we tested domain resolution via a custom DNS proxy on port 5553. Figure~\ref{fig:nslookup} illustrates \texttt{nslookup} queries for different subdomains (e.g., \texttt{www.xxx.ddns}, \texttt{ipv6.xxx.ddns}), each returning correct addresses or canonical names.

\noindent
Hence, the final DNS resolution step verifies that:
\begin{itemize}
    \item The IPFS hash is fetched from the Phicoin blockchain (eliminating centralized control).
    \item The proxy retrieves and parses the JSON file from IPFS.
    \item Each subdomain’s record is accurately mapped, whether it is an A, AAAA, CNAME, or MX record.
\end{itemize}

\paragraph{Summary of Findings}
This testing confirms that our DDNS system reliably supports domain creation and updates across IPFS and the Phicoin blockchain, then provides correct DNS responses via a local proxy. The \textbf{transaction hashes} in \lstinline|update_result| logs serve as irrevocable evidence of each record binding, while IPFS hashes guarantee data authenticity. Together, they form a robust foundation for a censorship-resistant, tamper-evident domain name service.

\subsection{Security Analysis}

\textbf{Identity and Trust Management}

\begin{itemize}
    \item \textbf{User Roles:}
    \begin{itemize}
        \item \textbf{pTLD Operators:} Control top-level domains (e.g., \texttt{.ddns}) through ownership of the corresponding blockchain assets~\cite{phicoin_whitepaper}.
        \item \textbf{Subdomain Owners:} Manage second-level domains under a pTLD, with rights transferred via blockchain transactions.
        \item \textbf{Visitors:} Users who perform domain name resolutions without needing Phicoin.
    \end{itemize}
    \item \textbf{Security Mechanisms:}
    \begin{itemize}
        \item \textbf{Private Keys:} Domain creation, modification, and deactivation require transactions signed with the owner's private key~\cite{phicoin_whitepaper}.
        \item \textbf{Immutable Records:} Domain records are bound to unique IPFS hashes, ensuring data integrity~\cite{benet2014ipfs}.
    \end{itemize}
\end{itemize}

\textbf{Confidentiality, Integrity, and Availability (CIA)}

\begin{itemize}
    \item \textbf{Confidentiality:} While the blockchain is transparent, sensitive data (like private keys) remain confidential~\cite{nakamoto2008bitcoin}. Domain ownership changes require private key signatures, ensuring only authorized modifications.
    \item \textbf{Integrity:} The immutability of blockchain records and IPFS's content-addressable storage ensure that domain records cannot be tampered with without detection~\cite{li2021bdns}.
    \item \textbf{Availability:} Decentralization eliminates single points of failure. Even if some nodes fail or are attacked, the system remains operational~\cite{duan2018dnsledger}.
\end{itemize}

\textbf{Resistance to Attacks}

\begin{itemize}
    \item \textbf{DNS Hijacking Prevention:}
    \begin{itemize}
        \item Local resolution using the blockchain and IPFS reduces reliance on external servers, mitigating hijacking risks~\cite{li2021bdns}.
        \item Cryptographic verification of data authenticity prevents attackers from injecting false records~\cite{yakubov2018blockchain}.
    \end{itemize}
    \item \textbf{DNS Cache Poisoning Prevention:}
    \begin{itemize}
        \item End-to-end data integrity verification ensures that only valid records are accepted~\cite{son2010hitchhiker}.
        \item The decentralized nature of the system reduces the effectiveness of poisoning attacks~\cite{li2021bdns}.
    \end{itemize}
    \item \textbf{Censorship Resistance:}
    \begin{itemize}
        \item No central authority controls the domain records, making it difficult for adversaries to censor or block domains~\cite{patsakis2020unravelling}.
        \item Users can run local nodes, accessing and sharing data without centralized intermediaries~\cite{ali2016blockstack}.
    \end{itemize}
\end{itemize}

\subsection{Trust Chain Analysis}
A fully decentralized domain name system (DDNS) must distribute trust among various components, each playing a distinct role in guaranteeing security, availability, and resistance to manipulation. Table~\ref{tab:trust_chain} summarizes these components and their responsibilities in the trust chain.

\begin{table}[H]
\centering
\caption{Key Components in the DDNS Trust Chain}
\label{tab:trust_chain}
\renewcommand{\arraystretch}{1.15}
\resizebox{0.48\textwidth}{!}{%
\begin{tabular}{p{3cm} p{4.5cm}}
\toprule
\textbf{Component} & \textbf{Description} \\ 
\midrule

\textbf{Phicoin Blockchain} 
& Serves as an immutable ledger for domain ownership and records. The PoW mechanism ensures tamper-evident transaction history and decentralized authority over updates. \\

\textbf{Miners / Nodes}
& Validate and propagate blocks across the network, competing for block rewards. Their collective effort secures the blockchain under economic incentives, preventing a single entity from rewriting history. \\

\textbf{IPFS}
& Provides distributed storage for domain configuration files (e.g., JSON records). Each record is referenced by a unique hash, making data tampering and centralized failures less likely. \\

\textbf{Domain Owner}
& Holds private keys to register, modify, or disable domain assets. Any change to the DNS binding must be cryptographically authorized, ensuring only rightful owners can update records. \\

\textbf{Resolvers}
& Retrieve the latest domain information by querying the blockchain for IPFS hashes, then fetch and parse the corresponding record files. They shield end-users from the complexity of blockchain interactions. \\

\textbf{End Users}
& Initiate domain queries through DNS proxies or resolvers. Rather than relying on a centralized DNS operator, they benefit from a censorship-resistant and fault-tolerant resolution process. \\

\bottomrule
\end{tabular}}
\end{table}

\noindent
\textbf{Key Observations:}
\begin{itemize}
    \item \emph{Distributed Governance:} No single party can unilaterally manipulate domain data; changes require valid signatures and block confirmations.
    \item \emph{Immutable Storage:} The blockchain ledger and IPFS content-addressed files together provide strong integrity guarantees.
    \item \emph{Economic Security:} Miners secure the chain under Proof of Work, disincentivizing malicious rewrites by making attacks computationally expensive.
    \item \emph{User-Centric Design:} Resolvers handle the blockchain-IPFS interaction so end-users can access domains with minimal friction.
\end{itemize}

By dispersing trust across a blockchain, a distributed storage layer, and cryptographically secured roles, this DDNS design circumvents single points of failure and fosters a resilient, transparent ecosystem for domain name resolution.

\subsection{Performance Analysis}
This section evaluates the system’s performance by focusing on three key aspects: \textbf{transaction throughput}, \textbf{DNS query speed}, and \textbf{scalability} under a Phicoin + IPFS architecture.

\paragraph{1) Blockchain Transaction Speed}
In our DDNS design, each domain operation (creation, modification, or deletion) corresponds to a blockchain transaction averaging \textbf{546 bytes}. Given Phicoin’s \textbf{4\,MB block size} and a typical 15-second block interval, the system can theoretically sustain up to:
\[
\frac{4\,\text{MB}}{546\,\text{bytes}} \div 15\,\text{s} \approx 512\,\text{transactions per second}.
\]
This capacity comfortably accommodates the rate at which DNS records are typically updated in real-world scenarios, minimizing the risk of congestion for domain-related transactions. Consequently, DNS record changes—such as adding an \texttt{A} or \texttt{AAAA} record—are broadcast and confirmed on-chain with minimal delay, maintaining high responsiveness for domain management tasks.

\paragraph{2) DNS Query Speed}
On the query side, the Phicoin client employs an in-memory database to cache recently accessed records, significantly reducing local lookup latency. When a domain is requested, the client first checks its memory cache. If a record is found, the lookup completes without additional network overhead, resulting in near-instant resolution. If not cached, the client retrieves the IPFS hash from the on-chain record, then fetches the corresponding JSON file from IPFS. Although this involves extra steps compared to a cache hit, preliminary tests suggest that, under moderate load, query performance remains acceptable for most DNS needs.  
To further optimize long-tail performance, \emph{future IPFS caching} mechanisms—such as local pinning or distributed gateway caching—are planned. This would allow popular domains or commonly accessed subdomains to be served from nearby nodes, lowering retrieval latency and improving scalability as the number of queries grows.

\paragraph{3) Scalability with Phicoin \& IPFS}
The proposed DDNS solution stores domain metadata (e.g., record types, addresses) in JSON files on IPFS, while only the IPFS hash and domain identifier reside on-chain. Since IPFS hashes are \emph{fixed in size}, broadcasting or updating domain references on the blockchain does not inflate transaction sizes despite the varying complexity of DNS configurations. This design provides two major benefits:
\begin{enumerate}
    \item \emph{Flexible JSON Configuration:} The system can seamlessly integrate or extend DNS protocols—like adding TLSA or SRV records—simply by updating the JSON structure. No additional overhead is introduced on-chain since the hash reference remains the same fixed length.
    \item \emph{Lightweight On-Chain Footprint:} As the number of domains grows, the blockchain only stores minimal references (IPFS hashes), rather than the entire DNS record data. IPFS, in turn, scales horizontally to handle large or complex records.
\end{enumerate}
By decoupling domain data from the ledger, the DDNS can adapt to diverse DNS record types, future protocol expansions, or complex configurations without overburdening the network.

\paragraph{Summary}
In conclusion, Phicoin’s \textbf{4\,MB} blocks and \textbf{546-byte} transactions enable up to \textbf{512 TPS}, ensuring ample throughput for domain-related operations. The \textbf{in-memory cache} in the Phicoin client boosts local DNS resolution speed, while integration with \textbf{IPFS} opens the door for further caching optimizations to reduce query latency. Meanwhile, using \textbf{fixed-size IPFS hashes} and JSON-based records preserves on-chain compactness and protocol extensibility, positioning the system to handle both immediate DNS requests and long-term growth in a decentralized manner.

\subsection{Feature Enhancements}

\begin{itemize}
    \item \textbf{Support for Additional DNS Protocols:}
    Extend domain templates to include records such as TLSA for DANE, enabling secure certificate verification without traditional Certificate Authorities (CAs) \cite{hoffman2012dns}.

    \item \textbf{User Interface Improvements:}
    Develop intuitive tools and graphical interfaces for domain registration and management, thereby lowering the entry barrier for non-technical users \cite{singla2018blockchain}.

    \item \textbf{Public DDNS Resolution Nodes:}
    Deploy public nodes compatible with traditional DNS to facilitate adoption and ease of use for end-users \cite{li2021bdns}.

    \item \textbf{Open and Extensible DDNS Protocol:}
    Evolve the DDNS framework into a fully open, specification-driven standard that any participant can implement on the Phicoin blockchain. Rather than limiting domain registration to a single pTLD (\texttt{.ddns}), users would be free to create and operate their own pTLDs, crafting unique namespaces and business models for domain services. Under this paradigm, each pTLD owner may define policies, set pricing, or introduce novel record types, while remaining interoperable with existing DDNS resolvers. By publishing a clear on-chain transaction format and record schema, the community can integrate emerging DNS features without modifying core blockchain logic. Critically, the IPFS hash reference—being a fixed size—ensures that registering or modifying complex domain records imposes minimal overhead on-chain. This openness fosters a diverse ecosystem where new naming solutions, specialized TLDs, and third-party resolvers can flourish without centralized gatekeeping.
\end{itemize}

\section{Conclusion}

This paper presents a secure, efficient, and low-cost decentralized DNS system that addresses several shortcomings of the traditional DNS. By leveraging blockchain and IPFS technologies, the DDNS system provides a censorship-resistant and tamper-proof domain name resolution service. It enhances the robustness of the internet infrastructure and offers a viable complementary solution to existing DNS systems.

\textbf{Future Applications and Potential:}

The DDNS system paves the way for more decentralized internet services, promoting security, privacy, and equal participation. Its scalable and flexible design makes it a strong solution for future internet infrastructure developments.
\section{Acknowledgements}

I sincerely thank Professor Ross Burke from UC Berkeley’s I School for his invaluable guidance and support throughout the course, which greatly helped shape this project.

Special thanks to Peter Trinh provided me a solar farm that I can use free electricity to run Phicoin’s seeder nodes, mining pools, and test mining algorithm.

\subsection*{Code Links}

The codebase for this project is available at:

\begin{itemize}
    \item GitHub Repository: \url{https://github.com/GY19A/ddns}
\end{itemize}

\end{document}